\documentclass[a4paper]{article}
\pdfoutput=1
\usepackage{hyperref}
\hypersetup{
  pdfinfo={
    Title={A penalized inference approach to stochastic block modelling of community structure in the Italian Parliament},
    Author={Mirko Signorelli and Ernst C. Wit}
  }
}
\usepackage{pdfpages}

\title{A penalized inference approach to stochastic block modelling of community structure in the Italian Parliament}
\author{Mirko Signorelli and Ernst C. Wit}
\date{}

\begin{document}

\maketitle

\noindent Citation info:\\
Signorelli, M. and Wit, E. C. (2017), A penalized inference approach to stochastic block modelling of community structure in the Italian Parliament. \textit{Journal of the Royal Statistical Society: Series C}. DOI: 10.1111/rssc.12234\\

\noindent The published version of this manuscript is available with Open Access from the website of the Journal of the Royal Statistical Society: Series C at

\noindent \href{http://onlinelibrary.wiley.com/doi/10.1111/rssc.12234/full}{http://onlinelibrary.wiley.com/doi/10.1111/rssc.12234/full}



\section*{Supplementary material}

# A penalized inference approach to stochastic block modelling of community structure in the Italian Parliament

Mirko Signorelli

*Leiden University Medical Center and University of Groningen, The Netherlands, and University of Padova, Italy*

and Ernst C. Wit

*University of Groningen, The Netherlands*

**Summary.** We analyse bill cosponsorship networks in the Italian Chamber of Deputies. In comparison with other parliaments, a distinguishing feature of the Chamber is the large number of political groups. Our analysis aims to infer the pattern of collaborations between these groups from data on bill cosponsorships. We propose an extension of stochastic block models for edge-valued graphs and derive measures of group productivity and of collaboration between political parties. As the model proposed encloses a large number of parameters, we pursue a penalized likelihood approach that enables us to infer a sparse reduced graph displaying collaborations between political parties.

*Keywords*: Adaptive lasso; Bill cosponsorship; Community structure; Network; Penalized likelihood; Stochastic block model

## 1. Introduction

The legislative process in modern democracies typically involves three fundamental steps: the proposal of a bill, a discussion on its contents and a final vote on it. Throughout this process, many interactions and collaborations can arise between different political actors, who join their efforts to support, change or oppose a proposed legislation. The analysis of these interactions can, then, provide insight into the features and the mode of operation of different parliaments, and on the way and the extent to which these interactions can influence the legislative process.

Two types of data are often considered in this context. The first is represented by bill cosponsorships networks (Fowler, 2006; Rocca and Sanchez, 2007; Parigi and Sartori, 2014). A parliamentarian can sponsor a bill individually, or cosponsor it together with other parliamentarians. In the latter case, bill cosponsorship implies a formal collaboration between its proponents, who officially state their agreement and support of the legislation proposed. The second is given by roll-call votes (Kirkland, 2014; Dal Maso *et al.*, 2014), in which parliamentarians express their final decision on a bill.

In this paper we study bill cosponsorship in the Italian Chamber of Deputies over the last four legislatures, covering the period 2001–2015. We represent bill cosponsorships by means of an undirected graph, where a weighted edge displays the number of bills that two deputies have cosponsored together. Compared with other parliaments, such as the American Congress or the German Bundestag, a distinguishing feature in the history of the Italian Parliament is the presence of a large number of political factions. Our primary aim is to infer a graph that summarizes collaborations within and between parties from the network of bill cosponsorships, whose actors are the deputies.

We tackle this issue by viewing edges $e_{ij}$ in the graph as a result of a Poisson process that explicitly depends on group memberships of nodes $i$ and $j$, as well as on their individual attributes. The model that we propose builds on the *stochastic block models* that are employed in social network analysis, which we review in Section 1.1. We resort to generalized linear models and derive measures of group relevance, and of collaboration or repulsion between groups. Finally, we propose a penalized inference approach for stochastic block models that enables us to reduce model complexity. We show that, with the use of penalized likelihood methods, a sparse reduced graph representing collaborations (and repulsions) between political parties can be obtained directly from the signs of the model parameters.

Our analysis demonstrates the evolution of the Italian political system from a highly polarized political arena, in which deputies base collaborations on their identification with left- or right-wing values, towards an increasingly fragmented parliament, where a rigid separation of parties into coalitions does not hold any more, and collaborations beyond the perimeter of coalitions have become possible.

### 1.1. Stochastic block models

Community membership can play an important role in shaping social interactions. Social networks are often featured by the presence of clusters of units that are strongly linked between themselves and weakly connected to individuals that fall outside their cluster, so that ignoring the preferential attachment of units based on community membership can lead to misleading interpretations of the determinants of network ties. Thus, cluster identification and assessment of the relationship between groups of nodes in a network have been active topics of research in the analysis of social networks.

Stochastic block models were first introduced as a modification of the $p_1$-class of models for unweighted digraphs that was proposed by Holland and Leinhardt (1981). If we denote by $X_{ij}$ a Bernoulli random variable that takes value 1 if an arrow from node $i$ to node $j$ is present, and is 0 otherwise, then the $p_1$-model assumes that pairs of edges or dyads $Y_{ij} = (X_{ij}, X_{ji})$ are stochastically independent and expresses the probability of observing the arrow $X_{ij}$ as a function of four parameters, representing the density of the graph $\theta$, the tendency of arrows to be reciprocated, $\rho$, expansiveness, $\alpha_i$, and popularity, $\beta_j$, of nodes $i$ and $j$. Fienberg and Wasserman (1981) considered a situation in which a partition of units into $p$ groups, also called *blocks*, is available, proposing a more parsimonious representation where $\alpha_i$ and $\beta_j$ are replaced by $p$ expansiveness group effects $\alpha_r$, such that $\alpha_i = \alpha_{i'}$ for every $i$ and $i'$ belonging to block $r$, and $p$ popularity group effects $\beta_s$.

The definition of a stochastic block model was proposed by Holland *et al.* (1983). According to their definition, a probability distribution for a graph defines a stochastic block model if the random variables $X_{ij}$ are independent, and the random vectors $X_{ij}$ and $X_{kl}$ are identically distributed if nodes $i$ and $k$ are members of the same block $r$, and $j$ and $l$ are in the same block $s$. Stochastic block models imply that nodes within a block are stochastically equivalent, in the

sense that, if nodes $i$ and $k$ belong to the same block $r$, any probability statement on the graph is left unchanged by interchanging them. Holland *et al.* (1983) criticized the model that was proposed by Fienberg and Wasserman (1981), deeming it too restrictive, and advocated that the parameters $\theta$, $\alpha_r$ and $\beta_s$ should be replaced by one parameter $\theta_{rs}$ for each pair of blocks $(r,s)$.

Later, Wang and Wong (1987) proposed a network model that retains the original formulation of the $p_1$-model with individual effects $\alpha_i$ and $\beta_j$ but also includes a set of block interaction parameters $\phi_{rs}$: one for each pair of blocks $(r,s)$.

Anderson *et al.* (1992) elaborated on the idea of stochastic block models, viewing them as

> 'a mapping of approximately equivalent actors into blocks or positions and a statement regarding the relations between the positions'.

They considered the $p_1$-class of models, and they proposed to represent relational ties between blocks of units by means of a reduced graph. They obtained such a graph setting a cut-off $c$ on the predicted probability of observing an arrow from nodes in group $r$ to nodes in group $s$, $\hat{\pi}_{rs}$, and drawing an arrow from $r$ to $s$ if $\hat{\pi}_{rs} > c$.

Stochastic block models have also been employed for community detection in networks, rather than to describe relationships between blocks of nodes that are known *a priori*. This type of block modelling aims to find clusters of highly interconnected nodes and it is referred to as *a posteriori* block modelling (Wasserman and Anderson, 1987; Nowicki and Snijders, 2001).

## 2. Bill cosponsorship in the Italian Parliament

The Italian Parliament is based on a bicameral system in which two separate assemblies, the Chamber of Deputies and the Senate, play similar roles in the legislative process. Legislations can be proposed by different actors (including deputies, senators, the government, regions and groups of electors); here, we focus on legislation proposed by deputies. Each bill can be proposed by a single deputy, or cosponsored by a group of deputies. In the second case, bill cosponsorship defines a symmetric relationship between deputies, who formally state their agreement on the content of the proposed legislation by cosponsoring it. Thus, cosponsorship can be taken as a measure of proximity or collaboration between deputies.

Bill cosponsorship can be represented as an undirected network where nodes represent parliamentarians, and the presence of an edge $e_{ij}$ indicates that parliamentarians $i$ and $j$ have cosponsored at least one legislation. We associate with each edge a weight equal to the number of bills that the two parliamentarians have sponsored together in a given time course (typically, one legislature).

In the Italian Chamber, each deputy is required to express their affiliation to one and only one parliamentary group, which typically corresponds to a political party or to a coalition of parties. As a consequence, membership of parliamentary groups generates a partition of deputies into political groups, which we use to assess the patterns of collaboration between political parties.

Data on bill cosponsorship in 27 parliamentary chambers of 20 European countries have been recently collected by Briatte (2016), who has created and published the corresponding cosponsorship networks. Here we consider the cosponsorship networks for the Italian Chamber of Deputies between the XIVth and the XVIIth legislatures (2001–2015) and we integrate these data with personal details on deputies retrieved from the Web site of the Chamber of Deputies (`http://dati.camera.it`).

The data that are analysed in the paper and the programs that were used to analyse them can be obtained from

`http://wileyonlinelibrary.com/journal/rss-datasets`

## 3. Poisson process model of bill cosponsorship

A graph is a pair $\mathcal{G} = (V, E)$, which consists of a set of nodes $V = \{1, \ldots, n\}$ connected by a set of edges $E \subseteq V \times V$. Edges represent relationships between nodes, and they can be directed or undirected, as well as weighted or unweighted. In bill cosponsorship networks, each node represents a parliamentarian and a weighted undirected edge between two parliamentarians displays the number of bills that they have cosponsored together. Thus, hereafter we consider the case of an undirected graph, where a discrete weight is associated with each edge. Such a graph can be conveniently represented by means of a symmetric adjacency matrix $Y$, where we set $y_{ij} = 0$ if deputies $i$ and $j$ are not connected, and $y_{ij}$ equal to the number of cosponsorships between deputies $i$ and $j$ otherwise. We assume absence of self-loops, i.e. $y_{ii} = 0$.

### 3.1. Data-generating process

We view the process of creation of edges in the graph as the result of a multivariate Poisson process in a given time course $T$. To wit, we can associate a Poisson process $N_{ij}(t)$ with rate $\lambda_{ij}$ with each pair of deputies $(i, j)$ in the graph. At the beginning of the legislature, i.e. $t = 0$, no cosponsorship has occurred yet, so $N_{ij}(0) = 0$. If after some time $t_1$ a first cosponsorship takes place between deputies $i$ and $j$, we set $N_{ij}(t_1) = 1$. If a second interaction occurs at $t_2$, we set $N_{ij}(t_2) = 2$, and so on. Thus, $N_{ij}(t)$ denotes the number of bill cosponsorships that have occurred between $i$ and $j$ at a given time point $t$. If we stop the process at $t = T$, the number of cosponsorships $N_{ij}(T)$ that are observed until $T$ between each pair $(i, j)$ of deputies is a realization from a Poisson distribution with mean $\mu_{ij} = \lambda_{ij}T$ and it defines a weighted graph, where $y_{ij} = N_{ij}(T)$.

Now, suppose that a partition $\mathcal{P}$ of deputies into $p$ groups or blocks is available, and that block membership determines the rates of each Poisson process, so that we can assume that the interaction rates $\lambda_{ij}$ are homogeneous within each pair of blocks $(r, s)$:

$$\lambda_{ij} = \zeta_{rs} \ \forall \, i \in \text{group } r, \ \forall \, j \in \text{group } s, \qquad r, s \in \{1, \ldots, p\}. \tag{1}$$

Under the assumption of independence between the univariate processes, equation (1) defines a stochastic block model, because $N_{ij}(t)$ and $N_{kj}(t)$ are independent, and they are also identically distributed if $i$ and $k$ belong to the same block. Our primary interest is to understand which groups are more active in the network, and how members from different groups interact with each other. Thus, we would like to decompose $\mu_{rs} = \zeta_{rs}T$ into a baseline parameter $\theta_0$ that controls the overall bill cosponsorship activity in the network, two main effects $\alpha_r$ and $\alpha_s$ that account for the relative importance (productivity or popularity) of political parties $r$ and $s$, and an interaction term $\phi_{rs}$ that accounts for collaboration (if positive), indifference (if null) or repulsion (if negative), between pairs of parties. Since a linear relationship between $\mu_{rs}$ and $\theta_0, \alpha_r, \alpha_s$ and $\phi_{rs}$ is impossible for the range $\mathbb{R}^+$ of $\mu_{rs}$, we consider a monotone transformation $g: \mathbb{R}^+ \to \mathbb{R}$ of $\mu_{rs}$ to be linear in the parameters, i.e.

$$g(\mu_{rs}) = \theta_0 + \alpha_r + \alpha_s + \phi_{rs}. \tag{2}$$

A convenient choice for $g$ is represented by the logarithm, but alternative choices for $g$ can be considered as well.

The stochastic block model in equation (2) implies stochastic equivalence of nodes within each block. As already noted by Wang and Wong (1987), stochastic equivalence is often an unrealistic and restrictive assumption. First, it is reasonable to imagine that deputies from the same party might behave differently. Furthermore, factors other than bill cosponsorship could

also play a role in the choice to cosponsor bills. Therefore, we extend model (2) to let the mean of each univariate process depend also on a set of node- or edge-specific covariates $\mathbf{x}_{ij}$, with an associated parameter vector $\boldsymbol{\beta}$:

$$
\begin{aligned}
y_{ij}|(i \in r, j \in s, \mathbf{x}_{ij}) &\sim \text{Poi}(\mu_{ij} = \lambda_{ij}T), \\
g(\mu_{ij}) &= \theta_0 + \alpha_r + \alpha_s + \phi_{rs} + \mathbf{x}_{ij}\boldsymbol{\beta}.
\end{aligned} \quad (3)
$$

Model (3) is not a proper stochastic block model, because it allows $\mu_{ij} \neq \mu_{kj}$ for two units $i$ and $k$ belonging to the same group $r$. Nevertheless, it retains its focus on the role that is played by blocks in shaping the network, including specific sets of parameters $\alpha_r$ for block relevance and $\phi_{rs}$ for interactions within and between blocks. Clearly, model (2) can be derived as a particular case of model (3) by setting $\boldsymbol{\beta} = \mathbf{0}$.

Model estimation can be performed by specifying a suitable generalized linear model (Nelder and Wedderburn, 1972; McCullagh and Nelder, 1989). We model the data-generating process in equation (3) with

$$
\log(\mu_{ij}) = \theta_0 + \sum_{r=1}^{p} \alpha_r D_r(i) + \sum_{r=1}^{p} \alpha_r D_r(j) + \sum_{r \leqslant s}^{p} \phi_{rs} D_{rs}(i, j) + \mathbf{x}_{ij}\boldsymbol{\beta}, \quad (4)
$$

where $D_r(i) = I(i \in r)$ and $D_{rs}(i, j) = I(i \in r, j \in s \vee i \in s, j \in r)$ for $r \leqslant s = 1, \ldots, p$ are dummy variables that indicate whether a unit $i$ belongs to group $r$, or whether the pair of nodes $(i, j)$ implies an interaction between blocks $r$ and $s$. However, model (4) is not identifiable without further constraints. Typically the way in which identifiability constraints are specified is not particularly important, as each parameterization is equivalent; however, as we shall be penalizing some parameters in later sections, the parameterization will be important. Thus, we introduce the following identifiability conditions:

$$
\sum_{r=1}^{p} \alpha_r = 0 \quad \text{and} \quad \sum_{s=1}^{p} \phi_{rs} = 0 \qquad \forall\, r = 1, \ldots, p, \quad (5)
$$

where for ease of notation we write $\phi_{sr} = \phi_{rs}$. If we incorporate these constraints in equation (4) by letting $\alpha_1 = -\Sigma_{r=2}^{p}\alpha_r$ and $\phi_{rr} = -\Sigma_{s \neq r}\phi_{rs}$, $\forall\, r = 1, \ldots, p$, the model can be rewritten as

$$
\log(\mu_{ij}) = \theta_0 + \sum_{r=2}^{p} \alpha_r T_r(i) + \sum_{r=2}^{p} \alpha_r T_r(j) + \sum_{r<s}^{p} \phi_{rs} T_{rs}(i, j) + \mathbf{x}_{ij}\boldsymbol{\beta}, \quad (6)
$$

where $T_r(i) = D_r(i) - D_1(i)$, $r \neq 1$, and $T_{rs}(i, j) = D_{rs}(i, j) - D_{rr}(i, j) - D_{ss}(i, j)$, $r \neq s$.

### 3.2. Extendibility

The model that we propose differs from traditional statistical models, where the outcome variable refers to a single statistical unit. An edge $e_{ij}$ involves, in fact, two statistical units, $i$ and $j$. This, in turn, implies that covariates that measure individual features ought to be transformed into edge attributes before they can be included in model (6). As an example, the sex, F or M, of two nodes gives rise to three possible edges: edges involving two males, MM, two females, FF, or one male and one female individual, FM. The ages of two individuals could be transformed into their absolute difference, or some other transformation such as their average, minimum or maximum.

The unusual nature of this model makes us examine its relevant invariance properties. Wit and McCullagh (2001) introduced the concept of extendibility of a statistical model, arguing that a sensible model is the model that, depending on the particular circumstances, can accommodate further treatments, or fewer covariate levels or changes of measurement scale than those actually observed. They advocated that invariance under selection of treatments, merging of covariate

levels and changes of measurement scale should be explicitly discussed when a new statistical model is introduced, and they showed that some commonly used models fail in this respect.

In our context, one could wonder whether it is sensible to require invariance with respect to group selection (introduction or elimination of a party), group merging (union of two existing parties) or changes of the measurement scale for $y_{ij}$. The answer to the first two points is strictly connected to what we consider to be a group: in the context of bill cosponsorship networks, each deputy joins a parliamentary group, so a block is a group of deputies who share similar political views and come together to promote the same political agenda. We therefore would like our model to retain its structure irrespectively of the fact that certain groups of individuals have been included or excluded from the analysis. However, if two parliamentary groups were to be merged this would produce a new political group, whose features would be different from those of either of the two original groups. For these reasons, we require model invariance under selection of groups, whereas we do not require invariance under group merging.

Invariance under selection of groups requires that, if one group—say the $p$th group—is excluded from model (6) and the new model

$$\log(\mu'_{ij}) = \theta'_0 + \sum_{r=2}^{p-1} \alpha'_r T_r(i) + \sum_{r=2}^{p-1} \alpha'_r T_r(j) + \sum_{r<s}^{p-1} \phi'_{rs} T_{rs}(i,j) + \mathbf{x}_{ij}\boldsymbol{\beta},$$

$$\text{subject to } \sum_{r=1}^{p-1} \alpha'_r = 0 \text{ and } \sum_{s=1}^{p-1} \phi'_{rs} = 0, \forall\, r=1,\ldots,p-1, \quad (7)$$

is considered, then it is possible to derive the parameters of model (7) as a function of the parameters of model (6). Indeed, this can be achieved by imposing $\mu'_{rs} = \mu_{rs}, r \leqslant s = 1,\ldots,p-1$ (selection requirement), and solving the resulting system of linear equations.

Finally, one might wonder whether it would be sensible to require invariance with respect to changes of measurement scale. Since the edge weights $y_{ij}$ are counts, it does not make sense to apply translations or dilatations to $y_{ij}$. However, we can consider changes of timescale and ask how this affects the block means $\mu_{rs}$. Consider a change of timescale from a system A with time expressed as $T_A$ and rates as $\zeta^A_{rs}$ to a system B with time $T_B$ and rates $\zeta^B_{rs}$. For example, system A could consider days and system B hours as time unit, so that $T_A = T_B/24$ and $\zeta^A_{rs} = 24\zeta^B_{rs}$. More generally, we can let $\zeta^A_{rs} = k\zeta^B_{rs}, k>0$. Since $T^A = k^{-1}T^B$, the block means $\mu_{rs}$ are not affected by the change of time system:

$$\mu^A_{rs} = T^A \zeta^A_{rs} = k^{-1} T^B k \zeta^B_{rs} = T^B \zeta^B_{rs} = \mu^B_{rs}.$$

This result implies that the parameters $\theta_0$, $\alpha_r$ and $\phi_{rs}$ in model (2) are left unchanged, so that the model is invariant with respect to changes of timescale measurement.

## 4. Inference

### 4.1. Parameter estimation

The parameter vector $\boldsymbol{\theta} = (\theta_0, \alpha_2, \ldots, \alpha_p, \phi_{12}, \phi_{13}, \ldots, \phi_{p-1,p}, \boldsymbol{\beta})$ that is associated with model (6) has dimension $q = \dim(\boldsymbol{\theta}) = p(p+1)/2 + \dim(\boldsymbol{\beta})$. In principle, it could be estimated with maximum likelihood. However, the number of model parameters $q$ increases quadratically with the number of blocks $p$. In such cases, maximum likelihood estimation could yield solutions with an extremely large number of parameters, making interpretation cumbersome. Instead, we advocate the use of penalized likelihood methods to achieve a parsimonious solution.

Besides enhancing model interpretability, penalized likelihood methods enable us to detect potentially sparse block-model-generating mechanisms. In stochastic block models, the block

interaction parameter $\phi_{rs}$ indicates an attraction ($\phi_{rs} > 0$) or repulsion ($\phi_{rs} < 0$) between the pair of blocks $(r,s)$, but it can also indicate indifference between some pairs of blocks—a situation that translates into $\phi_{rs} = 0$ in model (3). Whereas maximum likelihood is unlikely to produce model estimates $\hat{\phi}_{rs}$ that are exactly null, penalized likelihood is capable of distinguishing these cases of indifference by shrinking to 0 some of the block interaction parameters.

Since the introduction of the lasso (Tibshirani, 1996), penalized inference has become a popular choice for variable selection and the solution of high dimensional problems. Many methods in this field have been introduced (see Bühlmann and van de Geer (2011) and Fan and Li (2001) for an overview). In this paper we use the adaptive lasso (Zou, 2006), which is a weighted extension of the least absolute shrinkage and selection operator (the lasso) that was introduced by Tibshirani (1996), because it has good consistency properties.

The adaptive lasso aims for a sparse model solution by maximizing a penalized likelihood that incorporates the log-likelihood of the model, and a weighted $l_1$-penalty on the parameters that are included in the model. This penalty is multiplied by a tuning parameter $\delta \geq 0$, which determines the amount of regularization that is imposed on the parameters. The adaptive lasso problem for model (6) is

$$\max_{\boldsymbol{\theta}} \log\{L(\boldsymbol{\theta})\} - \delta \sum_{j=1}^{q} w_j |\theta_j|, \qquad (8)$$

where $L(\boldsymbol{\theta})$ denotes the likelihood of the model and $w_j$ is the weight that is associated with the $j$th element $\theta_j$ of $\boldsymbol{\theta}$. The tuning parameter $\delta$ is typically chosen either by cross-validation, or by minimizing a suitably defined information criterion. We discuss this issue in more detail in Section 4.2.

Denote by $\boldsymbol{\theta}^*$ a consistent estimator of $\boldsymbol{\theta}$ and by $N = n(n-1)/2$ the total number of pairs of nodes in the network. The attractive feature of the adaptive lasso is that if the weight vector is defined as $\mathbf{w} = 1/|\boldsymbol{\theta}^*|^{\gamma}$, and if $\delta/\sqrt{N} \to 0$ and $\delta N^{(\gamma-1)/2} \to \infty$, then the adaptive lasso estimator $\hat{\boldsymbol{\theta}}$ is consistent in variable selection (see theorem 4 in Zou (2006)).

The choice of the parameters that are subject to the $l_1$-penalty mostly depends on the role and the meaning that we associate with them. In our view, the block interaction parameter $\phi_{rs}$ expresses the presence of a collaboration or repulsion between deputies in parties $r$ and $s$ after we have accounted for both the overall density of the network, $\theta_0$, and the relevance of the groups, $\alpha_r$ and $\alpha_s$. To retain this interpretation, we do not penalize $\theta_0$ nor $\alpha_r$, $r = 1, \ldots, p$, i.e. we set $w_j = 0$ if $j \in \{1, \ldots, p\}$.

However, we aim to achieve some sparsity in the representation of relationships between groups by penalizing the $\phi_{rs}$-coefficients ($r \neq s$), as well as $\beta$. For the penalty weights, we compute the maximum likelihood estimate $\hat{\boldsymbol{\theta}}$ and set $w_j = 1/|\hat{\theta}_j|^{\gamma}$, with $\gamma = 2$, for $j > p$.

Because of the identifiability conditions in equation (5), the parameters $\phi_{rr}$ ($r \in \{1, \ldots, p\}$) that control interactions within each block do not explicitly appear in model (6) and, thus, they cannot be penalized. This implies that only the $p(p-1)/2$ interactions between different blocks can be penalized, whereas the $p$ parameters for within-block interactions are subsequently derived as $\hat{\phi}_{rr} = -\Sigma_{s \neq r} \hat{\phi}_{rs}$, $\forall r = 1, \ldots, p$. In practice, in real networks with community structure those parameters are typically strongly positive and, thus, unlikely to be shrunk to 0. For this reason, we believe that the parameterization in equation (6) represents the best compromise between identifiability and the need to penalize as many interaction terms as possible.

*4.2. Model selection*

In a penalized likelihood framework, the tuning parameter $\delta$ determines the amount of regularization that it is imposed on the parameters and, eventually, the level of sparsity of the solution.

Two main approaches are typically employed for the selection of an optimal tuning parameter $\delta^*$: cross-validation, or minimization of model information criteria. In the latter case, we seek

$$\delta^* = \arg\min_{\delta} (\mathcal{D}_\delta + a_m h_\delta), \tag{9}$$

where $\mathcal{D}_\delta$ denotes the deviance of the model, $m$ the number of observations and $h_\delta$ the dimensionality of the model. Various choices have been proposed for $a_m$. Alongside Akaike's information criterion AIC, which sets $a_m = 2$, and the Bayesian information criterion BIC, which takes $a_m = \log(m)$, recent proposals include the generalized information criterion GIC of Fan and Tang (2013), where $a_m = \log\{\log(m)\} \log(h_\delta)$, and the modified BIC MBIC of Chand (2012), where $a_m = \sqrt{m/h_\delta}$.

Here, we consider four simulations to assess the performance of these criteria in the selection of $\delta$. In each simulation, we generate a sequence of networks with increasing number of nodes $n = 50, 100, 150, \ldots, 500$, following the block model that is defined by equation (2). We set $\theta_0 = 0.7$ and draw $\alpha_r \in U(-0.3, 0.3)$, $r > 1$. Moreover, we set some $\phi_{rs}$-coefficients, $r \neq s$, equal to 0 and draw the remaining coefficients in such a way that $|\phi_{rs}| \sim U(c_{\min}, c_{\max})$, with $c_{\max} = 0.5$. Coefficients $\alpha_1$ and $\phi_{rr}$, $r = 1, \ldots, p$, are subsequently derived from equation (5). The simulations differ for the number $h$ of null $\phi_{rs}$-coefficients ($r \neq s$) and for the betamin condition ($|\phi_{rs}| \geq c_{\min}$) that is imposed on the non-null $\phi_{rs}$-coefficients; Table 1 in the on-line supplementary material summarizes the different settings in each simulation.

We perform model selection over a grid of 100 $\delta$-values. Each selection criterion leads to an optimal $\delta$ and corresponding model estimates. To compare the performance of each criterion in the selection of models that are capable of correctly distinguishing signals ($\phi_{rs} \neq 0$) and non-signals ($\phi_{rs} = 0$), we compute the accuracy of each solution, i.e.

$$\text{accuracy} = \frac{\text{true positives} + \text{true negatives}}{p(p-1)/2},$$

and we compare it with the maximum achievable accuracy for the set of 100 models that are considered. As shown in Fig. 1 of the supplementary material, every criterion quickly achieves the maximum accuracy when a dense model is considered (simulation A), but the accuracy of cross-validation, AIC and MBIC is often lower when sparser models are considered (simulations B and C), or when signal detection is complicated by the imposition of a milder betamin condition (simulation D). Overall, BIC and GIC outperform the competing methods and, thus, they appear to be the best information criteria for variable selection.

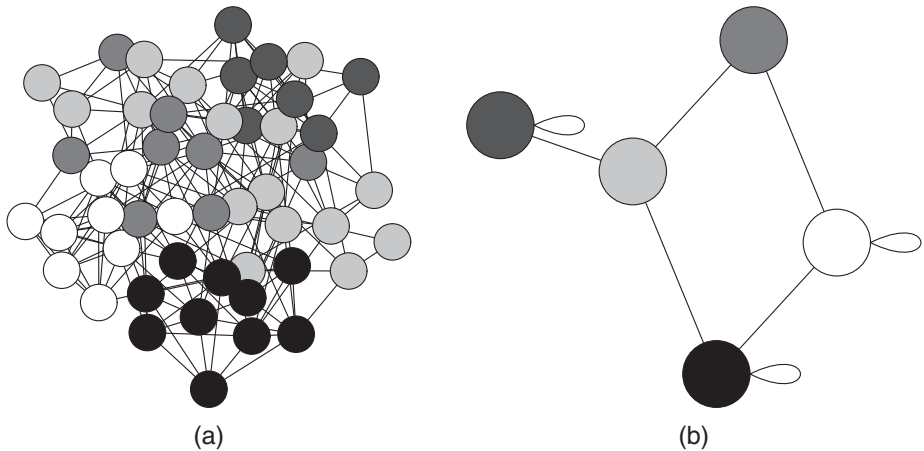

**Fig. 1.** (a) An unweighted graph with 50 nodes, partitioned into five groups and (b) a simplified representation of relationships between groups: ☐, set 1; ▨, set 2; ▧, set 3; ▪, set 4; ■, set 5

*4.3. The reduced graph*

A focal aspect of stochastic block models is the description of the relationships between blocks of individuals. Anderson *et al.* (1992) proposed to represent relational ties between blocks of units by means of a reduced graph, whose nodes are the blocks. The idea behind this reduced graph is quite simple: summarize the original graph by visualizing relationships between blocks directly, to achieve a simpler and clearer representation.

As an example, consider the graph in Fig. 1(a). Three groups of nodes (sets 1, 4 and 5) appear to be featured by a strong internal connectivity; besides, nodes within each group tend to be preferentially linked to nodes belonging to one or two other groups; for example, it appears that nodes in set 3 tend to prefer nodes in sets 1 and 2 to nodes in sets 4 and 5. On the basis of similar observations, we can attempt to draw a reduced graph that summarizes our intuition: the graph in Fig. 1(b) provides an example.

Different strategies to derive a reduced graph from a statistical model can be considered. Anderson *et al.* (1992) obtained such a graph setting a cut-off $c$ on the predicted probability of observing an arrow from nodes in a group $r$ to nodes in a group $s$, $\hat{\pi}_{rs}$, and drawing an arrow from $r$ to $s$ if $\hat{\pi}_{rs} > c$. The resulting reduced graph links blocks that are highly connected, but edges therein do not necessarily display attraction between groups. For example, nodes in a group $r$ could have overall higher degrees: if this is so, block $r$ would be connected to any group, just as a result of the high average degree of nodes in the block. Moreover, their approach cannot be easily generalized to edge-valued graphs.

Therefore, we propose an alternative strategy to derive a reduced graph displaying collaborations between parties, which is based on the parameter estimates $\hat{\phi}_{rs}$ in model (6) rather than on $\hat{\mu}_{rs}$ (or $\hat{\pi}_{rs}$). By doing so, we control for the average degree of blocks $r$ and $s$, as well as for the effect of individual covariates. Since an estimate $\hat{\phi}_{rs} > 0$ entails evidence of collaboration between deputies in parties $r$ and $s$, we draw an edge between blocks $r$ and $s$ if $\hat{\phi}_{rs} > 0$.

Furthermore, it is also possible to derive a reduced graph that displays repulsions by connecting blocks such that $\hat{\phi}_{rs} < 0$. In an unpenalized likelihood framework, however, such a reduced graph is uninteresting, as it is simply the complement of the reduced graph of collaborations. Instead, as discussed in Section 4.1, penalized inference enables us to distinguish collaborations and repulsions from situations of indifference between parties. In a penalized likelihood setting, then, the reduced graph of repulsions is not just the complement of the reduced graph of collaborations, but it becomes an interesting outcome of model estimation that can highlight those pairs of parties whose members avoid working with each other.

## 5. Analysis of bill cosponsorship networks of the Italian Chamber of Deputies

We consider now the networks representing bill cosponsorship in the Italian Chamber of Deputies, which we have described in Section 2. We focus our attention on the cosponsorship networks of the four legislatures XIV–XVII, covering the period 2001–2015. During this period, the number of parliamentary groups ranged from 8 (legislatures XIV and XVI) to 10 (XVII) and 13 (legislature XV); in each legislature, a mixed group has always been present, gathering deputies from small political groups with different political orientation, which did not meet the requirements (defined in the Chamber's regulations) for the creation of a parliamentary group.

We study the dependence between bill cosponsorship and parliamentary groups, controlling for some individual attributes of the deputies. In particular, we consider gender, education level (undergraduate *versus* graduate), age, seniority and the electoral constituency of each deputy. Gender can give rise to edges involving two male, MM, two female, FF, and a female and a male, FM, deputies; we take MM as reference. Likewise, we take interactions between two un-

dergraduate deputies, UU, as reference and introduce dummies for graduate–undergraduate, GU, and graduate–graduate, GG, interactions. We distinguish senior deputies, S, who had already been parliamentarians before their election in a given legislature, from junior deputies, J, who were first experiencing being deputies. We set interactions between junior deputies, JJ, as reference mode and introduce two dummies for junior–senior, JS, and senior–senior, SS, interactions. Furthermore, we consider the age difference of the two deputies. We take Lombardia as reference electoral constituency, and we introduce 20 fixed effects for the remaining constituencies (19 regions plus the constituency for electors living abroad). We also consider a dummy indicating whether two deputies have been elected in the same constituency.

A commonly observed feature of social networks is the presence of triadic effects. For binary graphs, these triadic effects correspond to the fact that the probability of observing an edge between two individuals increases with the number of common neighbours that they share. This idea is the basis of exponential random graph models for binary graphs (Frank and Strauss, 1986), whose estimation relies on Markov chain Monte Carlo simulation techniques (Snijders, 2002) and is typically unfeasible for networks featuring more than a few hundred nodes. Extensions of exponential random graph models for edge-valued graphs have been recently proposed (Desmarais and Cranmer, 2012; Krivitsky, 2012), but the estimation of the transitivity effect for large networks remains an open issue. To account for triadic effects, we consider for each pair of nodes $(i, j)$ the statistic $TR_{ij} = \Sigma_{k \neq i,j} y_{ik} y_{jk}$, whose value increases with the number of shared cosponsors, as well as with the frequency of cosponsorships undertaken with them. We include $TR_{ij}$ in model (6) and estimate its parameter with a penalized pseudolikelihood approach. We remark that the performance of penalized pseudolikelihood in the estimation of the transitivity term of exponential random graph models has not been investigated yet, and the inclusion of $TR_{ij}$ in the model should be regarded just as an attempt to account for transitivity effects on bill cosponsorship.

For each legislature, we estimate model (6) with the adaptive lasso, using BIC to select the tuning parameter $\delta$. Table 1 shows the estimates of $\theta_0$ and $\beta$ (except for the regional effects, which are reported in Table 2 of the on-line supplementary material). The estimate of the intercept $\theta_0$ is lower for legislatures XV and XVII, coherently with the fact that the networks for those legislatures refer to shorter timeframes (less than 3 years *versus* the 5 years of legislatures XIV

**Table 1.** Effect of individual attributes on bill cosponsorship†

| *Covariate* | *Results for the following legislatures:* | | | |
|---|---|---|---|---|
| | XIV | XV | XVI | XVII |
| Intercept $\theta_0$ | −2.693 | −3.184 | −2.767 | −3.598 |
| Female–male FM | 0.139 | 0.155 | 0.208 | 0.211 |
| Female–female FF | 0.604 | 0.714 | 0.689 | 0.642 |
| Graduate–undergraduate GU | 0.155 | 0.000 | 0.011 | 0.000 |
| Graduate–graduate GG | 0.158 | 0.000 | 0.000 | −0.157 |
| Same electoral constituency | 0.527 | 0.516 | 0.537 | 0.535 |
| Junior–senior JS | −0.045 | 0.043 | 0.000 | 0.231 |
| Senior–senior SS | −0.004 | 0.127 | 0.176 | 0.571 |
| Age difference | −0.020 | 0.000 | −0.061 | −0.040 |
| Transitivity | 0.189 | 0.131 | 0.058 | 0.067 |

†The table displays the estimates of $\theta_0$ (unpenalized) and $\beta$ (penalized) in model (6) for the following legislatures: XIV (2001–2006), XV (2006–2008), XVI (2008–2013) and XVII (2013–2015).

and XVI). Bill cosponsorships turn out to be more frequent between female deputies (FF) and, in general, they are more likely to take place if at least one of the sponsors is female (FM). The effect of education, instead, is not stable over time. The positive estimates that are associated with pairs of deputies who were elected in the same electoral constituency clearly point out that deputies tend to collaborate on the basis of geographic proximity. Whereas in legislature XIV junior deputies were slightly more productive than their senior colleagues, from legislature XV onwards cosponsorships involve more senior than junior deputies. Moreover, cosponsorships are more frequent between deputies of similar age. Finally, we find evidence of transitivity effects. The effects that are associated with each constituency (Table 2 of the supplementary material) are mostly shrunk to 0 and they do not point to any peculiar temporal pattern.

The pattern of interactions between political parties can be reconstructed by inspecting the reduced graphs of collaborations in Fig. 2, where an edge displays collaborations ($\hat{\phi}_{rs} > 0$) between two parliamentary groups, a self-loop indicates that there is a tendency of deputies to cosponsor with deputies from the same parliamentary group and node size is proportional to the relative frequency of cosponsorship, $\hat{\alpha}_r$, of deputies in each group. Conversely, the reduced graphs representing repulsions ($\hat{\phi}_{rs} < 0$) between parties are shown in Fig. 2 of the on-line supplementary material.

The first, interesting, conclusion is that cosponsorships during legislatures XIV and XV reflected collaborations within each party, and between parties that belonged to the same political coalition. In fact, both legislatures featured strong competition between two coalitions, one of which (the right wing in the first case, and the left wing in the latter) held the majority in the parliament and could, thus, govern on its own. This situation seems to have generated a strong ideological polarization, which is evident from the pattern of collaborations (and repulsions) between the parliamentary groups.

The division of the Chamber into two coalitions ended with legislature XVI, as a centrist party (the Unione di Centro, UDC) that was not part of any coalition entered the Chamber. For 3 years, the majority was held by the right-wing coalition, whereas the UDC and the left-wing coalition were in opposition. 3 years later, a group of right-wing deputies formed the Futuro e Libertà per l'Italia party, FLI, a new political group that abandoned the right-wing coalition and entered a centrist coalition with the UDC. 1 year later, the right-wing government resigned and a coalition government, supported by a heterogeneous coalition of parties, took its place. Besides cosponsorships within each parliamentary group, our model detects collaborations between the main right-wing party (the Popolo della Libertà party, PDL) and FLI, between two opposition parties (Partito Democratico, PD, and UDC) and between a left-wing party (the Italia dei Valori party, IDV) and a right-wing group (the Popolo e Territorio party, P&T). It is also interesting to consider the reduced graph displaying repulsions: most of the edges therein indicate (not surprisingly) the absence of collaborations between parties from different coalitions, but also between the UDC and FLI, which allied towards the end of the legislature. In short, cosponsorships in this legislature seem to reflect mostly the division between the right-wing majority (FLI, the Lega Nord party, LN, PDL and P&T) and the opposition (PD, IDV, UDC) of the first half of the legislature, despite the fact that the analysis considers cosponsorships over the whole legislature span. A possible explanation for this result is that cosponsorship events are more likely to take place in the first years of each legislature: as a matter of fact, owing to the long time that is typically necessary for a bill of parliamentary initiative to be discussed and approved, a bill that is proposed towards the end of the legislature is extremely unlikely to be approved, and this can in turn discourage deputies from proposing bills in the last years of their mandate.

The fragmentation in the composition of the Chamber has become even stronger in the current legislature (XVII). Since none of the four coalitions now represented in the Parliament (left

wing, right wing, the centrist Scelta Civica, SC, and the Movimento 5 Stelle, M5S) could form a government alone, alliances between parties belonging to different coalitions had to be sought, giving rise to heterogeneous parliamentary majorities. In this case, the reduced graph in Fig. 2 shows that, besides self-loops accounting for a tendency towards within-group cosponsorship, deputies from different right-wing parties collaborate with each other. Moreover, deputies from the centrist party SC collaborate with deputies belonging to the Centro Democratico party, CD, a left-wing party which is ideologically alike the SC but belongs to a different political coalition. Further collaborations are detected between two left-wing parties (PD and the Sinistra Ecología

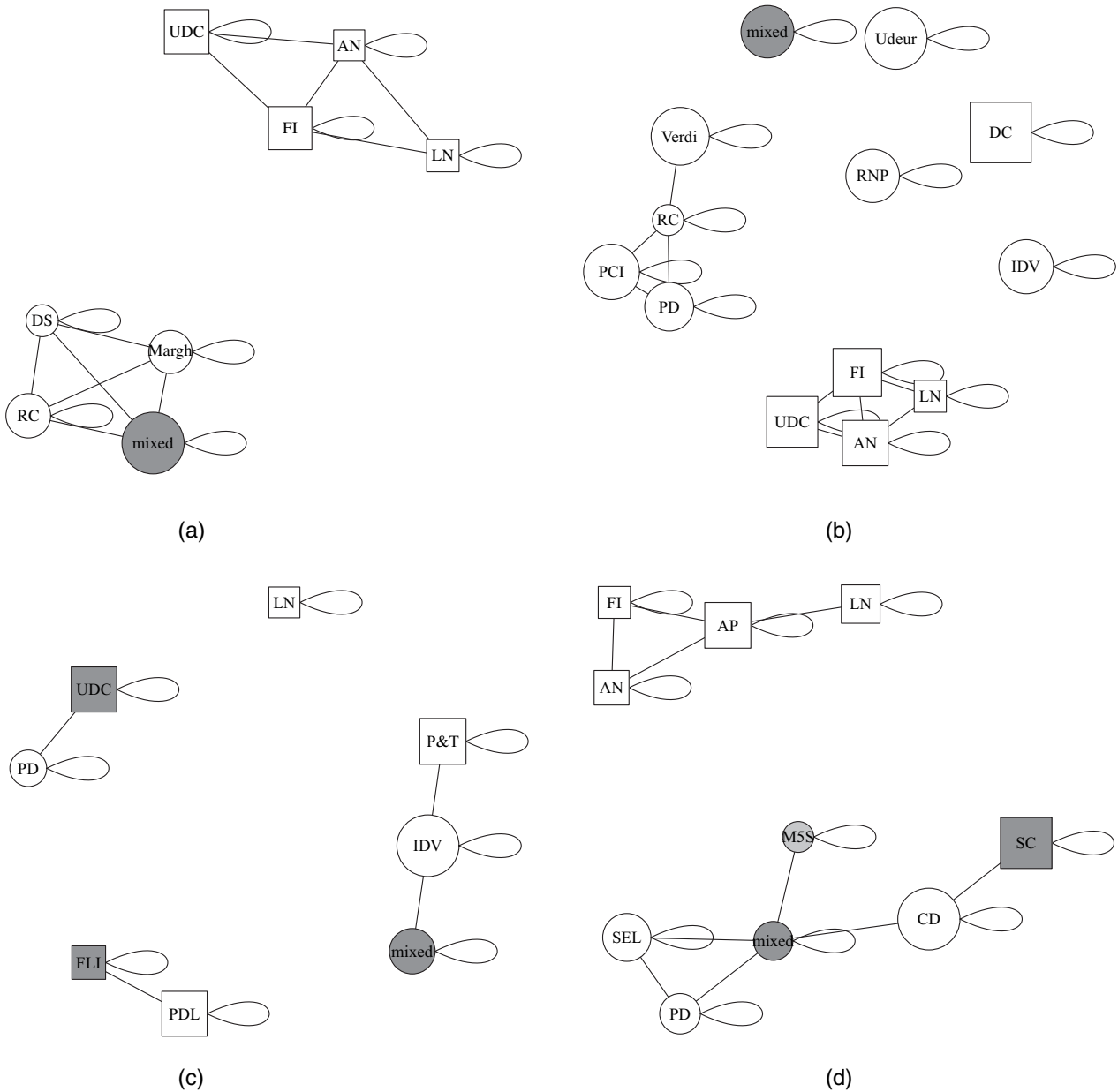

**Fig. 2.** Reduced graphs representing collaborations between parliamentary groups based on bill cosponsorship (the graphs display collaborations based on model 6 (i.e. $\hat{\phi}_{rs} > 0$); □, right-wing parliamentary groups; ○, left-wing groups; ■, centrist groups; ●, the mixed group; ◐, M5S) (node size is proportional to the productivity of each parliamentary group, $\hat{\alpha}_r$): (a) legislature XIV (2001–2006); (b) legislature XV (2006–2008); (c) legislature XVI (2008–2013); (d) legislature XVII (2013–2015)

Libertà party, SEL) and between the mixed group and various parties. Apart from a collaboration with the mixed group, deputies from M5S do not seem to collaborate with any other party.

In short, our analysis of bill cosponsorship networks indicates the evolution from a highly polarized political arena, in which deputies based collaborations on their identification with left- or right-wing values, towards an increasingly fragmented parliament, where a rigid separation of political groups into coalitions does not seem to hold any more, and collaborations beyond the perimeter of coalitions have become possible. One of the drivers of this change is probably a change of electoral law in 2005, which made it more difficult for coalitions of parties to obtain a majority in the Senate. This resulted in the premature end of legislature XV and in less stable parliamentary majorities in legislatures XVI and XVII. Our analysis of bill cosponsorships suggests that also the pattern of collaborations between parliamentarians was affected, inducing deputies to collaborate more frequently with deputies from different political coalitions.

## 6. Conclusion and discussion

Community affiliation can deeply affect social behaviour and the formation of relationships between individuals. In social network analysis, stochastic block models represent a popular approach to account for the effect of community membership on the creation of ties and to assess community structures.

In this paper, we have developed an extended stochastic block model for the analysis of bill cosponsorships in the Italian Parliament. This model retains the focus on relationships between pairs of blocks that characterize pure stochastic block models by including parameters for group productivity, $\alpha_r$, and interactions between pairs of groups, $\phi_{rs}$, but it also allows heterogeneity of units within a block. Because the number of parameters increases quadratically with the number of groups, we advocate the use of a penalized estimation approach to select a parsimonious model that displays relevant collaborations and repulsions between pairs of blocks. We represent these preferential relationships by means of reduced graphs displaying the relationships that exist between blocks.

Our analysis of bill cosponsorship in the Italian Chamber of Deputies from 2001 to 2015 demonstrates the evolution from a political system that was strongly polarized into a left- and a right-wing coalition, in which bill cosponsorship took place almost exclusively between deputies belonging to the same coalition, towards an increasingly fragmented political arena, with more than two coalitions of parties and in which collaborations beyond the perimeter of coalitions are now possible.

Although here we have considered networks where edges are undirected and weighted, with weights in the set of natural numbers, the models that we propose can be easily generalized in two directions. Directed edges can be handled by introducing a reciprocity term and a further set of nodal effects to distinguish sender and receiver nodes. As an example, model (3) can be adapted as follows:

$$y_{ij}|(i \in r, j \in s) \sim \text{Poi}(\mu_{ij}),$$
$$\log(\mu_{ij}) = \theta_0 + \rho + \alpha_r + \gamma_s + \phi_{rs} + x_{ij}\boldsymbol{\beta},$$

where $\alpha_r$ measures the productivity of group $r$ (which the sender node $i$ belongs to), $\gamma_s$ the popularity of group $s$ (which the receiver node $j$ belongs to) and $\rho$ the tendency to reciprocate arrows. Here a positive $\phi_{rs}$ denotes attraction or repulsion from nodes in group $r$ towards nodes in group $s$, and, consequently, $\phi_{rs} \neq \phi_{sr}$.

Moreover, the use of generalized linear models enables us to extend model (3) easily beyond Poisson processes. For example, if the network is unweighted (i.e. $y_{ij} \in \{0, 1\}$) it suffices to replace

the Poisson with a Bernoulli distribution, and the log-link with a logit or a probit link function; if a weighted network with weights in the set of real numbers is at hand, the Poisson distribution can be replaced with any continuous distribution, and the identity function becomes a natural choice for $g$.

We remark that our data analysis relies on bill cosponsorship networks that are aggregated over the span of each legislature (Briatte, 2016). This does not allow us to take into account possible changes in membership of parliamentary groups within a legislature, a practice—known as *trasformismo*—that is quite frequent in the Italian Parliament. For this reason, we have relied on the group memberships of each deputy as reported on the Web site of the Italian Chamber of Deputies (`http://dati.camera.it`). In principle, our model is capable of handling this situation. If, for example, deputy $i$ has been member of party $q$ for a timespan equal to $t_1$ and of party $r$ for $t_2$, the number of bills that they have cosponsored with deputy $j \in s$ is still a Poisson process:

$$N_{ij}(t_1 + t_2) = N_{ij}(t_1) + N_{ij}(t_2) \sim \text{Poi}(\lambda_{qs} t_1 + \lambda_{rs} t_2).$$

Thus, availability of data disaggregated over time would allow us to cope with these changes in party membership, providing a more realistic account of this phenomenon. Furthermore, this would also entitle us to model directly the interaction rates $\lambda_{ij}$ between deputies, which (as we pointed out in our comment on the results for legislature XVI) is unlikely to be constant across the legislature (because of both procedural issues and the changing political environment). In particular, it would make it possible to verify the hypothesis that most cosponsorships take place at the beginning of the legislature.

## Acknowledgements

*Supporting information*

Additional 'supporting information' may be found in the on-line version of this article:

'Web-based supporting materials for: "A penalized inference approach to stochastic blockmodelling of community structure in the Italian Parliament"'.

# Web-based supporting materials for: "A penalized inference approach to stochastic blockmodelling of community structure in the Italian Parliament" by Mirko Signorelli and Ernst C. Wit

Mirko Signorelli[1,2,3] and Ernst C. Wit[1]

[1] *Johann Bernoulli Institute for Mathematics and Computer Science, University of Groningen (NL)*
[2] *Department of Statistical Sciences, University of Padova (IT)*
[3] *Department of Medical Statistics and Bioinformatics, Leiden University Medical Centre (NL)*

Table 1: **An overview of Simulations A-D.** In Simulation A, we consider a dense model (i.e., with high dimensionality $h$) with a moderate betamin condition imposed on the non-null $\phi_{rs}$ coefficients ($|\phi_{rs}| \geq c_{\min}$). We progressively increase the sparsity of the model in Simulations B and C. In Simulation D we consider a model with medium sparsity level (like the one in Simulation B), but we make signal detection harder by imposing a milder betamin condition.

| Simulation | # ($\phi_{rs} = 0$) | $h$ | Betamin condition |
|---|---|---|---|
| A | 10 | 45 (dense) | $c_{\min} = 0.2$ (moderate) |
| B | 20 | 35 (medium) | $c_{\min} = 0.2$ (moderate) |
| C | 30 | 25 (sparse) | $c_{\min} = 0.2$ (moderate) |
| D | 20 | 35 (medium) | $c_{\min} = 0.1$ (mild) |

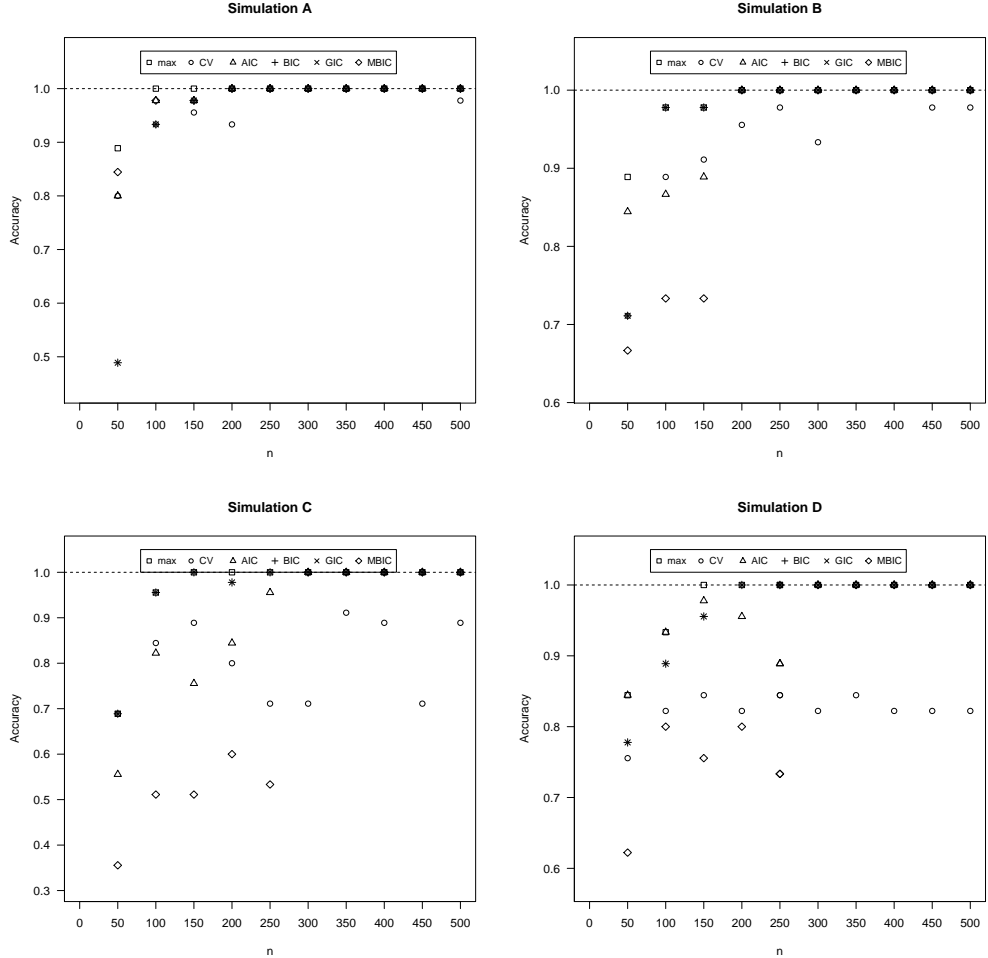

Figure 1: **Results of Simulations A-D.** Comparison of the accuracy of models chosen by 10-fold cross-validation (CV), Akaike's Information Criterion (AIC), Bayesian Information Criterion (BIC), the Generalized Information Criterion (GIC) of Fan and Tang (2013) and the modified BIC (MBIC) of Chand (2012) with the maximum achievable accuracy (MAX). Every criterion quickly achieves the maximum accuracy in Simulation A, where we consider a model with few null $\phi_{rs}$. In Simulations B, C and D, instead, BIC and GIC outperform CV, AIC and MBIC: this is particularly apparent when a sparser model is considered (Simulation C), or when signal detection is made harder by the imposition of a milder betamin condition (Simulation D).

2

Table 2: **Parameter estimates of regional effects** (reference mode: Lombardia). Note that most of the effects are shrunk to zero and that, overall, the effects are not constant over time.

| Costituency | Legislature | | | |
|---|---|---|---|---|
| | XIV | XV | XVI | XVII |
| Abruzzo | 0 | 0 | 0.240 | 0 |
| Valle d'Aosta | 0.750 | -0.262 | -0.274 | 0.373 |
| Basilicata | 0.146 | 0 | 0 | 0.313 |
| Calabria | -0.058 | 0 | 0.302 | 0 |
| Campania | 0 | 0 | 0.020 | -0.289 |
| Emilia Romagna | 0 | 0 | 0.103 | -0.173 |
| Friuli Venezia Giulia | 0 | 0 | 0 | 0 |
| Lazio | 0.113 | 0 | 0 | 0 |
| Liguria | -0.303 | 0 | 0 | -0.358 |
| Marche | 0.029 | 0 | -0.185 | 0 |
| Molise | -0.426 | 0.382 | 0.531 | 0 |
| Piemonte | 0.115 | 0 | -0.016 | 0 |
| Puglia | 0 | 0 | 0.221 | 0 |
| Sardegna | 0 | 0 | 0 | 0 |
| Sicilia | 0 | 0 | 0.132 | -0.240 |
| Toscana | 0 | 0.098 | 0.019 | -0.283 |
| Trentino Alto Adige | -0.028 | 0 | 0.332 | -0.260 |
| Umbria | 0 | -0.167 | 0.398 | 0 |
| Veneto | 0.126 | 0 | -0.281 | 0 |
| Residents abroad | - | 0.390 | 0.705 | 0 |

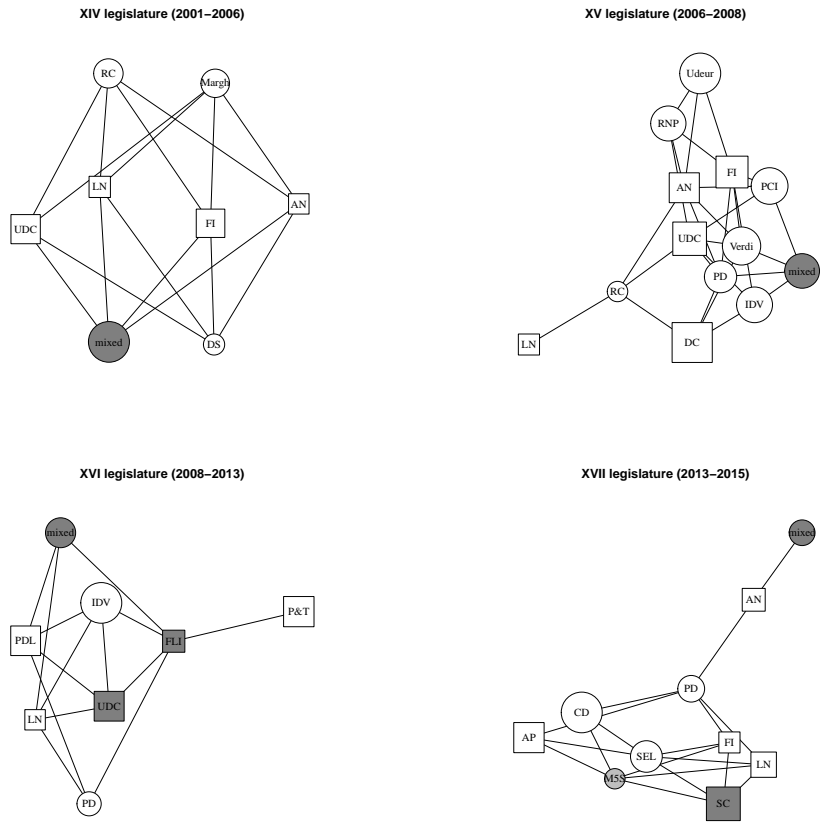

Figure 2: **Reduced graphs representing repulsions between parliamentary groups based on bill cosponsorship.** The graphs display repulsions (i.e., $\hat{\phi}_{rs} < 0$) between parliamentary groups. White squares denote right-wing parliamentary groups, white circles left-wing groups and darkgrey squares centrist groups. A darkgrey circle denotes the mixed group, whereas a lightgrey circle the Movimento 5 Stelle. Node size is proportional to the productivity of each parliamentary group ($\hat{\alpha}_r$).



\end{document}